# Nanobubble-driven superfast diffusion dynamics of Brownian particles

**Authors:** Xuewen Fu[*], Bin Chen, Jau Tang[*], Ahmed H. Zewail[†]

**Affiliations:**

Physical Biology Center for Ultrafast Science and Technology, Arthur Amos Noyes Laboratory of Chemical Physics, California Institute of Technology, Pasadena, CA 91125, USA

[*]Corresponding authors: xuewen@caltech.edu; jautang@caltech.edu

[†]Deceased, August 2016.

**Dynamics of active or self-propulsive Brownian particles in nonequilibrium status[1-3], has recently attracted great interest in many fields including biological entities[4,5] and artificial micro/nanoscopic motors[6]. Understanding of their dynamics can provide insight into the statistical properties of biological and physical systems far from equilibrium[7,8]. Generally, active Brownian particles can involve either translational or rotational motion. Here, we report the translational dynamics of photon-activated gold nanoparticles (NPs) in liquid cell imaged by four-dimensional electron microscopy (4D-EM). Under excitation of femtosecond (fs)-laser pulses, we observed that those Brownian NPs exhibit a superfast diffusive behavior with a diffusion constant four to five orders of magnitude greater than that in absence of laser excitation. The measured diffusion constant was found to follow a power-law dependence on the fs-laser fluence. Such superfast diffusion is induced by strong random driving forces arising from rapid nucleation, expansion and collapse of photoinduced nanobubbles (NBs) near the NP surface. In contrast, the motion of the NPs exhibit superfast ballistic translation at a short time scale down to nanoseconds (ns). Combining with physical model simulation, this study reveals a NB-propulsion mechanism for the self-propulsive motion, providing physical insights for better design of light-activated artificial micro/nanomotors.**

Back in 1827, using optical microscope the botanist Robert Brown first observed the jittery motion of small suspended particles and found that each moving step of the particle was independent of the previous one[9]. This random motion of the particles, namely, the well-known Brownian motion nowadays, was first interpreted by Einstein in 1905 as the amplification of thermal fluctuations: a Brownian particle receives momentum from the surrounding molecule collisions, but its movement is dissipated by the viscosity of the liquid[10]. For this passive Brownian particle in thermal equilibrium, its motion is diffusive at long times ($t \gg \tau_p$, $\tau_p$ is the momentum relaxation time)[10,11] but ballistic at short times ($t \ll \tau_p$)[12-14].
2

Over the past decade, active Brownian motion has attracted increasing interest due to the great potential of using micro/nanomachinery in biomedical applications[6]. For a Brownian object, an external driving force could drive it out of equilibrium to realize the autonomous motion[15]. Notable examples of such out-of-equilibrium systems range from the biological entities such as bacteria[4] and motile cells[5] to artificially inorganic micro/nanomotors[16-18]. They present the potential as autonomous agents for delivering nanoscopic objects to targeted positions, such as in non-invasive microsurgery[19] and drug delivery[20,21], where the interplay between the external activations and the internal fluctuations from the nature of the systems lead to a complex behavior of their dynamics. So far, a number of different propulsion mechanisms have been proposed for the artificially self-propelled micro/nanomotors. Most of these mechanisms are based on a scheme that the micro/nanomotors act as an engine and generate a propulsive force owing to the following reasons including magnetic field[22], electric field[23], special flow field induced by a gradient of osmotic pressure (self-electrophoresis) or temperature[24-27] and catalytic decomposition[28]. However, the diffusion process or the directional motion of the micro/nanomotors by these propulsion mechanisms is relatively slow, where the average speed is usually below several μm/s. For the phenomenon of superfast self-propulsive motion, on the other hand, is still far from being understood. Once achieved, it would provide the perspective of using micro/nanomachinery for efficient nanoscopic objects delivery in biomedicine.

In this report, we unravel a type of nanobubble (NB)-propulsion mechanism for superfast diffusion dynamics of photon-activated Brownian particles in liquid by 4D-EM. The solution containing spherical gold NPs (~80 nm in diameter) was sealed between two electron-transparent, 20 nm thick silicon nitride membranes with the liquid thickness of ~300 nm. Details of the preparation protocol of the liquid cell are described in the Methods and are also available elsewhere[29-31]. The liquid cell was then integrated to our 4D-EM for dynamical measurement (Fig. 1). In 4D-EM, green fs-laser pulses were utilized to trigger the dynamics while precisely timed electron pulses were used to image the particle motion (see Methods). For the measurement at long times, repetitive fs-laser pulses (wavelength: 520 nm; pulse duration: 350 fs) were used to activate the NP, while a train of electron pulses (repetition rate of 1 kHz) were presented to trace the trajectory of the NP motion. At short time scale, single-pulse imaging methodology was used to capture the transient morphologies of the NP activated by a single fs-laser pulse, where the images were acquired by a precisely timed electron pulse at specific delays. Details of the procedure are also available in the previous studies from this group[31-34].

We traced the translational dynamics of a single photon-activated gold NP at long time scale by stroboscopic electron pulses. A set of typical snapshots of the NP translational motion under repetitive fs-laser pulses excitation (fluence of 2.3 mJ/cm$^2$, repetition rate of 1 kHz) at different times are presented in Fig. 2a. Prior to excitation (0 s), the NP inside the thin layer solution sandwiched in the liquid cell was weakly bound near the substrate surface by weak electrostatic interaction[35,36]. Upon excitation, the NP was activated to move randomly due to the rapid nucleation, expansion and collapse of the steam NBs visible near the particle surface (see Fig. 2a). Note that the number, size and nucleation position of the NBs were independent at different times. Fig. 2b displays two typical trajectories of the NP under repetitive laser pulses excitation with different fluences (2.0 and 2.3 mJ/cm$^2$). These trajectories show that the particle translates in a manner of random walk and the range of its displacements (follow Gaussian distribution, data are not shown here) increases with the fluence, indicating the faster translation of the particle at a higher fluence.

To understand the statistical properties of the translational dynamics of the photon-activated gold NP, its mean square displacements (MSDs) under different laser fluences (1.6 - 3.0 mJ/cm$^2$, repetition rate: 1 kHz) are presented in Fig. 3a. All the measured MSDs almost show a linear relation with time, i.e., $MSD \propto t$, following a substantial increase in the slope with increasing fluence. These linear MSDs demonstrate that the photon-activated gold NP exhibits a diffusive behavior, which is similar to the conventional passive Brownian motion. However, the diffusion constant in our observations exhibits a power-law dependence on the laser fluence as $D \propto (J - J_c)^{2.2}$ for $J \geq J_c$ (see the fit curve in Fig. 3b), and its value ($1.2 \sim 36 \times 10^3$ nm²/s) is four to five orders of magnitude greater than that of the conventional Brownian motion of gold NPs without photon activation[35]. The parameter $J_c$ is the minimum laser fluence (threshold) required to raise water above its boiling temperature (400 K) to evaporate the water molecules near the gold NP surface. The retrieved threshold value of the fluence from the fit is 1.25 mJ/cm², which agrees well with the experimental observation that below a fluence of ~1.3 mJ/cm² the laser heating of the NP is insufficient to generate steam NBs. Moreover, at a fixed laser fluence ($J$ = 3.2 mJ/cm²), the diffusion of the gold NP becomes faster with increasing the laser repetition rate (see the MSDs in Fig. 3c), and the diffusion constant is proportional to the repetition rate (see Fig. 3d). From these results, an intuitive mechanism for the superfast diffusion of the photon-activated NP under laser excitation emerges. Due to the strong local photothermal effect as a result of the surface plasmon enhanced optical absorption of the gold NP at the laser wavelength, the particle is heated up in hundreds of picoseconds[32]. This results in a raised temperature over the boiling point (~400 K) of the neighboring water molecules. The water steam begins to nucleate as NBs on the NP surface, then the NBs expand, collapse and induce a random driving force on the NP to propel its motion. Unlike the random force arising from the fluid molecule collisions in the passive Brownian motion, it is the random force arising from the photoinduced NBs near the NP surface that dominates the superfast diffusion.

Using single-pulse imaging mode of 4D-EM, we further studied the transient dynamics of the photon-activated gold NPs at ns time scale. Figs. 4 presents the transient translation process of a gold NP induced by a single fs-laser pulse excitation with a fluence of 7.8 mJ/cm². The first and second columns show the typical single-pulse images of the NP before the laser pulse and at specific delays (25, 40 and 60 ns; single-pulse images of more delay points are not shown here), respectively; while their corresponding difference images are shown in the third column. The dash red and blue circles in the difference images indicate the initial position and the position of the NP at specific delays, respectively. The displacement of the NP increases with the delay time. Considering the ballistic feature of the Brownian motion at short time scale[12-14], the displacement evolution of the NP with delay time determined by different single-pulse imaging measurements was plotted in Fig. 4b. The displacement increases very slow in the first 25 ns and then grows fast from 25 to 60 ns (nearly linear increase), followed by a very slow increase to a saturation level after about 80 ns. This result demonstrates that upon the fs-laser pulse excitation the NP was accelerated and obtained a momentum in 20~30 ns due the photoinduced steam NBs near the particle surface, then its velocity and translation were damped in hundreds of nanoseconds by the surrounding drag, including both the liquid friction and the interaction from the substrate. The velocity of the NP (from 20 to 60 ns) was estimated to be ~ 6 m/s, which is three orders of magnitude faster than that of the conventional passive Brownian motion in the ballistic regime[14]. Therefore, the nature of the photon-activated Brownian motion is similar to the passive one at the short time range (both are ballistic), but the former one has a much higher velocity due to the strong impulsive driving force induced by the photoinduced NBs.

Theoretical simulation was performed to unravel the underlying mechanism for the observed superfast diffusion of the photon-activated gold NPs at both long and short time scales. Based on Einstein's theory, the one-dimensional translational dynamics of an active Brownian particle can be generally reformulated in terms of a Langevin equation[37] as

$$\frac{d^2}{dt^2}x + \gamma \frac{d}{dt}x = A_{th}(t) + A_{ext}(t) \qquad (1)$$

where $x$ is the particle displacement, $\gamma$ is the damping factor due to surrounding friction force, $A_{th}(t)$ is the acceleration induced by the random force arising from the liquid molecule collisions, and $A_{ext}(t)$ is the acceleration executed by external activations. For the photon-activated gold NPs in our experiment, it is reasonable to consider $A_{ext}(t) >> A_{th}(t)$.

For the dynamics at the long time scale, we proposed a simple physical model to analyze the MSD directly from the Langevin equation, with the assumption that $A_{ext}(t)$ consists of a comb of very short rectangular-shaped impulses that arise from the NBs induced by the repetitive laser pulses excitation with a duration of $\tau_0$ and a time interval of $T_p$ (repetition rate is $1/T_p$). For the $k$-th impulse, $A_{ext}(t) = A_k(0)\left(H(t - kT_p) - H(t - \tau_0 - kT_p)\right)$, where $H(t)$ is a Heaviside step function. The Langevin equation can be thus recast as

$$\frac{d^2}{dt^2}x + \gamma \frac{d}{dt}x = \sum_{k=0} A_k(0)\left(H(t - kT_p) - H(t - \tau_0 - kT_p)\right) \tag{2}$$

For simplicity, we assume the initial value $x(0)=0$ and the velocity $u_0 = \dot{x}(0) \neq 0$. Since there is neither correlation between the initial velocity and the impulsive forces, nor correlation between the directions of the impulsive forces in different excitation circles, one has $\langle u_0 A_k(0)\rangle = 0$ and $\langle A_k(0) A_j(0)\rangle = \delta_{kj} A_{ave}^2$, where $A_{ave}$ (assumed to depend on the laser fluence) represents an average value of the comb of random impulses. Assuming the damping constant is independent of liquid temperature and resolving the above Langevin equation (see Supplementary Text), we obtained a key formula for the MSD in the long time limit of $t >> 1/\gamma$ as

$$MSD \approx \frac{t}{T_p} \frac{A_{ave}^2 \tau_0^2}{\gamma^2} \tag{3}$$

Similarly, for a two-dimensional case, $MSD \approx 2\frac{t}{T_p}\frac{A_{ave}^2 \tau_0^2}{\gamma^2}$, which is linear with respect to time $t$ and is in good agreement with our experimental results (Figs. 3a and 3c). In analogy with Einstein's linear law for a conventional two-dimensional Brownian motion ($MSD = 4Dt$)[10], one can get an effective diffusion constant as $D = \frac{A_{ave}^2 \tau_0^2}{2T_p \gamma^2}$, which has a quadratic dependence on the strength of the average impulsive acceleration. As $A_{ave}$ is proportional to the laser fluence $J - J_c$ ($J_c$ is the threshold fluence required to raise the temperature to cause evaporation of the water molecules near the gold NP surface), the deduced effective diffusion constant should also follow a quadratic relation with the laser fluence, i.e. $D \propto (J - J_c)^2$ for $J \geq J_c$. Our experimental observation, $D \propto (J - J_c)^{2.2}$ (see Fig. 3b), agrees well with the simulation. The small discrepancy of the exponent indicates slight temperature dependence of the liquid viscosity (related to the damping constant in eq. (2)), which slightly decreases with the temperature of the surrounding liquid and steam nanobubbles. Furthermore, the predicted diffusion constant linearly increases with the impulse repetition rate $1/T_p$ at a given fluence, which is in line with the experimental result (see Fig. 3d). These agreements validate our hypothesis and the physical model of the NB-driving mechanism.

For the dynamics of the NP at short time scale induced by a single fs-laser pulse, the impulsive acceleration is given for simplicity by a rectangular-shaped impulse function with a time duration of $\tau_0$, namely, $A_{ext}(t) = A(0)(H(t) - H(t - \tau_0))$, where $A_{ext}(t) \gg A_{th}(t)$. Solving the Langevin equation (1) by substituting this external impulsive acceleration $A_{ext}(t)$, the analytical solution for $x(t)$ was obtained as

$$x(t) = \left\{-\frac{A(0)}{\gamma}\left[\frac{e^{-\gamma(t-\tau_0)}-1}{\gamma}+(t-\tau_0)\right]\right\} \cdot H(t-\tau_0) + \frac{A(0)}{\gamma^2}\left(e^{-\gamma t}-1\right)+\frac{A(0)t}{\gamma} \quad (4)$$

where the Heaviside step function $H(t-\tau_0)=1$ for $0 \leq t \leq \tau_0$, and $H(t-\tau_0)=0$ for $\tau_0 < t$. Therefore, for $0 \leq t \leq \tau_0$, one has $x(t) \approx \frac{A(0)}{2}t^2$, suggesting that in such a short time range the particle is still under acceleration; for $\tau_0 < t$, the acceleration period ends and due to the damping from the surrounding drag the particle displacement levels off, following eq. (4). This prediction captures well the single-pulse-imaging result in our experiment (see the fit curve by eq. (4) in Fig. 4b). The magnitude of the damping constant $\gamma$, which reflects the effective damping effect, was retrieved to be ~ 0.05 ns$^{-1}$. The acceleration period $\tau_0$ due to the NB driving effect was determined to be ~ 35 ns, which coincides with the lifetime range of the fs-pulse-induced steam NBs around the gold NPs in water[38], further validating the hypothesis of our model. Therefore, the combined experimental results (Figs. 3 and 4) and model simulations at both long and short times elucidate the NB-driving mechanism for the superfast diffusion of the photon-activated gold NPs. These findings would inspire ideas for the design of light-activated artificial micro/nanomotors, for example, Janus micro/nanostrcutrues with high propulsion efficiency.

Our results reveal a type of high efficient self-propulsive mechanism of gold NPs under photon excitation. Due to strong NB-driving forces, the active particles show a superfast diffusion with the diffusion constant being four to five orders of magnitude greater than that of the passive Brownian motion. This result provides an insight into the fundamental dynamical behavior of nanoscale self-propulsive motion, and may find applications in the design of light-activated nano/microscopic swimmers and machines. Because of the high spatiotemporal resolution of the liquid-cell 4D-EM, it would also permit the ability to study the dynamical behaviors of other nanoscopic or microscopic systems out-of-equilibrium, such as biological entities and active matter in their native environments.

**Acknowledgments**

This work was supported by the Air Force Office of Scientific Research Grant FA9550-11-1-0055S for research conducted in The Gordon and Betty Moore Center for Physical Biology at California Institute of Technology. We wish to thank J. S. Baskin, M. T. Hassan and H. Li for help on setting the femtosecond laser system.

**Author contributions**

A. H. Z. and X. W. F. conceived the research project. X. W. F. and B. C. did the measurement and the data analysis.  J. T. did theoretical analysis and model simulation. All the authors wrote the manuscript.

**Figure Legends**

**Figure 1. Schematic of photon-activated gold NP diffusion in liquid imaged by 4D-EM.** For the time-resolved imaging at long times, stroboscopic electron pulses are continuously present to image the NP diffusion under repetitive fs-laser pulses excitation. At short time scale (single-pulse imaging mode), a fs-laser pulse enables the photon-activated NP motion, while an precisely timed electron pulse is used to image the change of the transient NP morphologies at given time delays .

**Figure 2. Typical snapshots and trajectories of photon-activated gold NP diffusion in liquid. a**, Typical snapshots of a gold NP diffusion under 1 kHz laser pulses (fluence of 2.3 mJ/cm$^2$) excitation at the different times. The NP was driven to move by rapid nucleation, expansion and collapse of the photoinduced NBs near the particle surface (see the circles with white contrast). **b**, Two typical trajectories of the gold NP diffusion at different laser fluences of 2.0 and 2.3 mJ/cm$^2$, respectively. The NP moves in a manner of random walk.

**Figure 3. Laser fluence and repetition rate dependence of the gold NP diffusion dynamics**. **a**, MSDs of the gold NP diffusion under different laser fluences (repetition rate of 1.0 kHz). **b**, Variation of the diffusion constant of the photon-activated NP as a function of laser fluence, which follows a power-law dependence with a retrieved threshold fluence of $J_c \sim 1.25$ mJ/cm$^2$. Inset: schematic of superfast diffusion (random walk) of a NP driven by the nucleation, expansion and collapse of photoinduced NBs. **c**, MSDs of a gold NP diffusion under a fixed laser fluence of 3.2 mJ/cm$^2$ with different repetition rates. **d**, Variation of the diffusion constant as a function of the laser pulse repetition rate, which follows a linear dependence.

**Figure 4. Single-pulse imaging of a photon-activated gold NP in the ballistic regime. a**, Single-pulse images of the photon-activated NP at three specific delay times. The images in the first and second columns are the states of the NP before the laser pulse and at specific delays of 25, 40 and 60 ns after the laser pulse (fluence of 7.8 mJ/cm$^2$), respectively. The third column shows the corresponding difference images obtained by subtracting the single-pulse image before the laser pulse from that at the short delay time. The dash red and blue circles indicate the NP positions before the laser pulse and at short delays after the laser pulse, respectively. **b**, Evolution of the displacement of the NP as a function of delay time with a physical model fit. The inset shows the image of the measured gold NP.

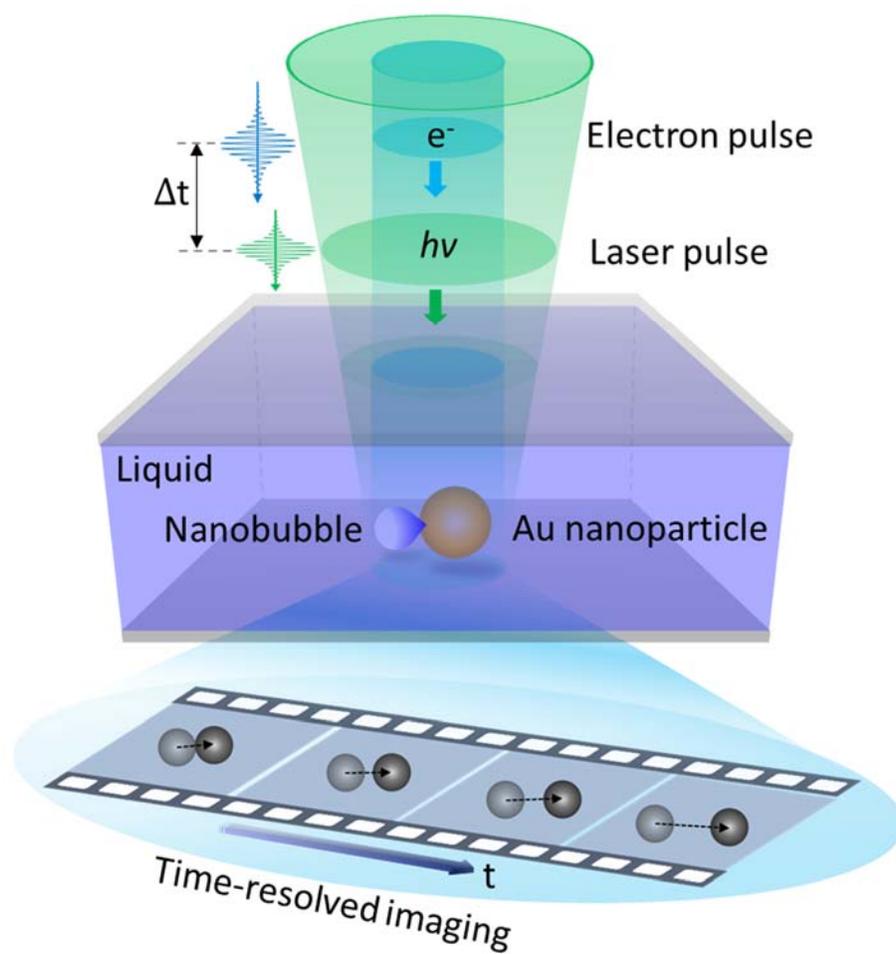

**Figure 1 (Fu *et al.*)**

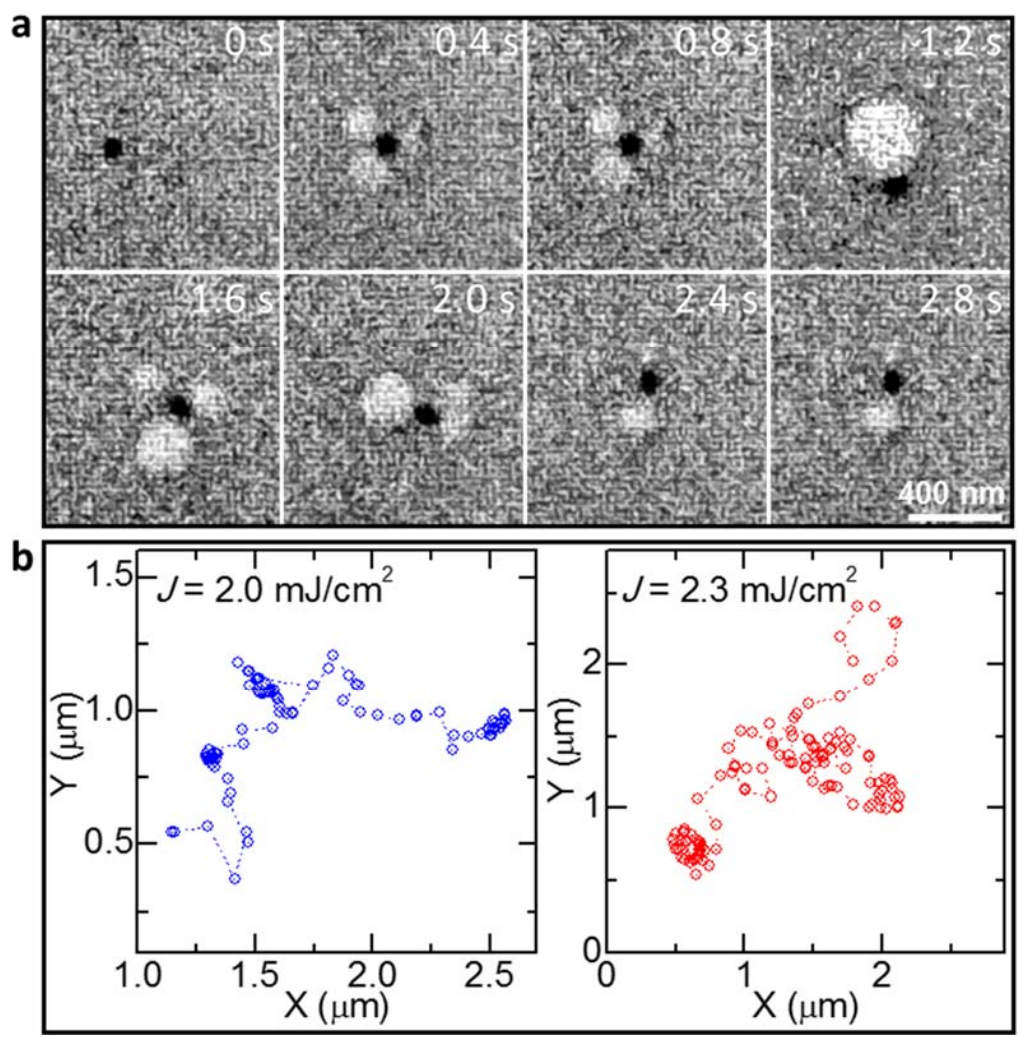

**Figure 2 (Fu *et al*.)**

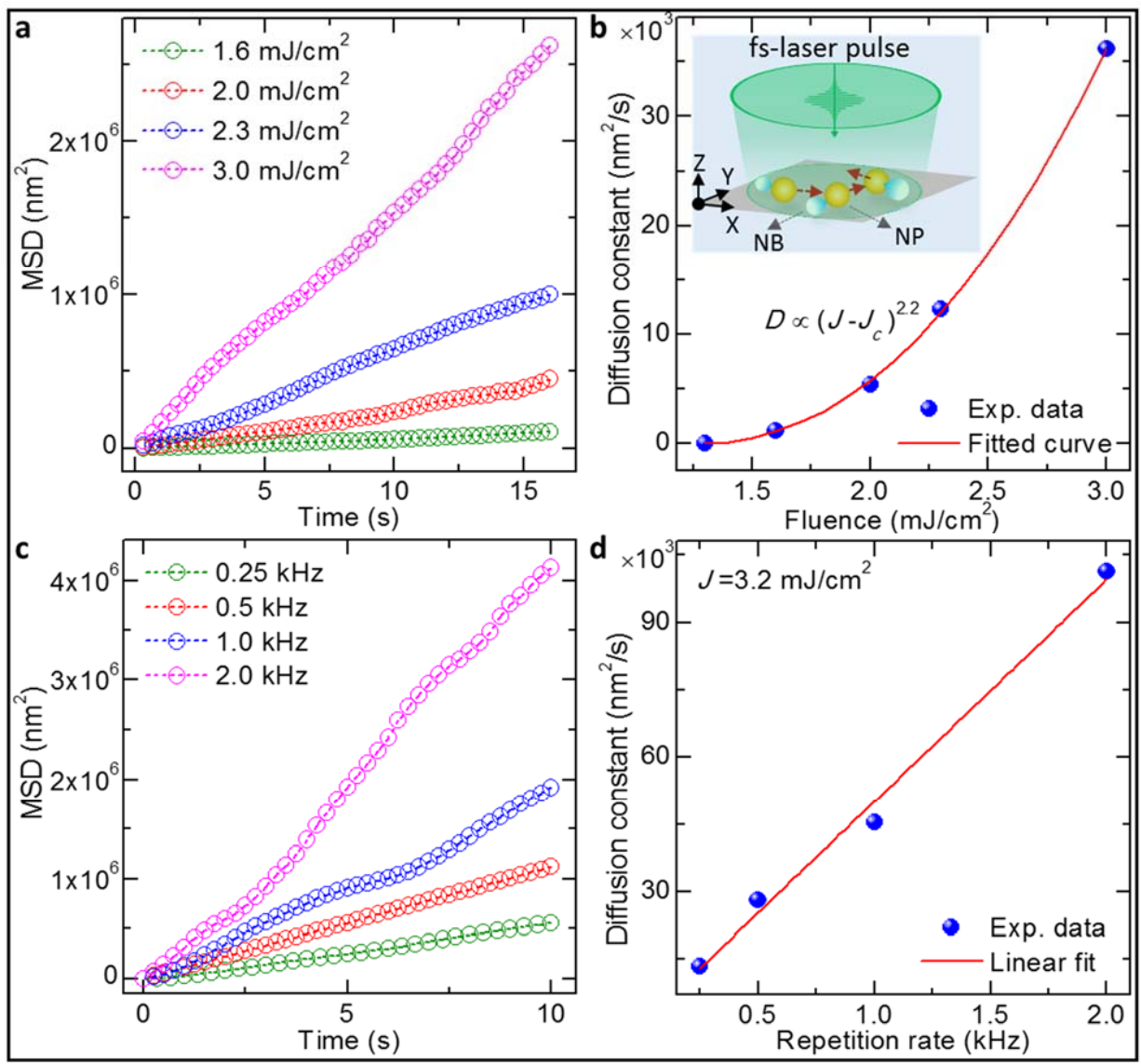

**Figure 3 (Fu *et al*.)**

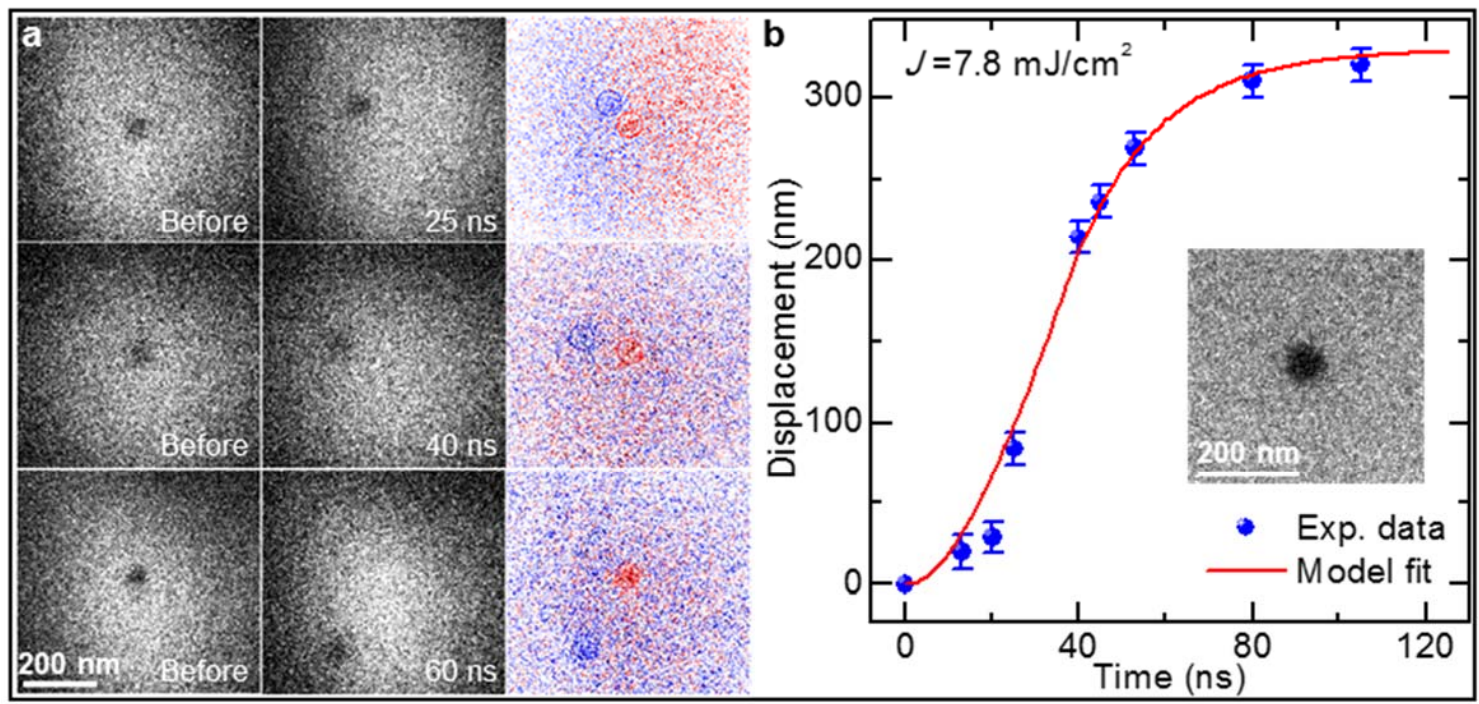

**Figure 4 (Fu *et al.*)**

# Supplementary Information for

# Nanobubble-driven superfast diffusion dynamics of Brownian particles

**Authors:** Xuewen Fu[*], Bin Chen, Jau Tang[*], Ahmed H. Zewail[†]

**Affiliations:**

Physical Biology Center for Ultrafast Science and Technology, Arthur Amos Noyes Laboratory of Chemical Physics, California Institute of Technology, Pasadena, CA 91125, USA

[*]Corresponding authors:  xuewen@caltech.edu; jautang@caltech.edu

[†]Deceased, August 2016.

**This PDF file includes:**

      Methods
      Supplementary Text
      References

# Methods

**Preparation of Liquid cell**

Standard transmission electron microscopy (TEM) $Si_3N_4$ windows ($250 \times 250$ μm$^2$) on 200-μm-thick silicon frame (customized from Norcada Inc., circular shape that fits the standard TEM holder) with a thin layer of pre-prepared gold film (~200 nm) as a spacer (customized from Norcada Inc.) were used in this work. The gold film spacer was deposited at the edge of the silicon frame, which would not block the electron transparent $Si_3N_4$ window. The ultrathin low-stress $Si_3N_4$ membrane, deposited by low-pressure chemical vapor deposition, is ~20 nm thick, which guarantees the spatial resolution of the liquid cell to be in nanometer range.

The protocol for preparation of liquid cell is as follows. The $Si_3N_4$ membrane was rinsed with acetone, isopropyl alcohol, and deionized water followed by plasma cleaning ($P \sim 3.5$ W, $t \sim 30$ s). A drop of ~5 μL aqueous solution with spherical gold nanoparticles (NPs) stabilized with citrate ligand (~80 nm in diameter, purchased from Nanocomposix Inc.) was loaded on the bottom chip with spacer, and another top chip without spacer was put on the surface of the liquid. Due to the tension of the liquid, the top $Si_3N_4$ window would rotate freely and align well with that of the bottom one. Then these two well-aligned $Si_3N_4$ window chips were clamped by a tweezer, and the superfluous liquid was absorbed by a small piece of filter paper from the side. A thin layer of liquid was sandwiched between the two chips with its thickness nearly equivalent to the spacer thickness. An epoxy adhesive (Ted Pella Inc.) was immediately used to seal the side of the liquid cell. After the sealing was dried, the liquid cell was loaded in our four-dimensional electron microscopy (4D-EM) for the measurements.

During the measurement, we checked each liquid cell by tilting it to different view angles and excluded the bad one with gas bubbles or apparent bulges. For the good liquid cell filled with liquid, there may be a little bulge when put in the high vacuum of TEM. Nevertheless, compared to the big window area ($250 \times 250$ μm$^2$), the observed photon-activated superfast diffusion of the gold NPs only occurs in a small area of $< 2 \times 2$ μm$^2$ (see Fig. 2b in the main text), the possible existence of a small bulge should not significantly influence the observed diffusion dynamics of the NPs in such a small area shown in this work.

**Time-resolved imaging measurements**

The measurements were performed in our 4D-EM instrument (UEM-1 at California Institute of Technology) with the integration of liquid cell. Femtosecond (fs) infrared laser pulses (1040 nm, 350 fs pulse duration) were used to generate the visible pump fs-laser pulses (520 nm) via second harmonic generation and the green fs-pulses were directed to the liquid cell sample inside the microscope to activate the particle dynamics. Precisely synchronized ultraviolet (UV) nanosecond (ns)-laser pulses (266 nm, 10 ns pulse duration) were directed to the photocathode inside the TEM to generate ultrafast electron pulses, which were accelerated to 120 keV to probe the particle dynamics through the Si$_3$N$_4$ window. A digital delay generator was used to synchronize the ns-electron-probe pulse and the fs-laser-excitation pulse and to control the time delay between them. For the time-resolved imaging measurement at long times, the sample in the liquid cell was excited by repetitive green fs-laser pulses, and stroboscopic electron-pulses (repetition rate of 1 kHz) were present to capture the dynamics. While for the time-resolved imaging measurement at short time scale, i.e. the single-pulse-imaging mode, a single green fs-laser pulse was used to trigger the dynamics and a precisely timed ns-electron pulse was used to probe the subsequent transient dynamics at specific delays. The time resolution of our 4D-EM instrument is determined to be ~10 ns[1]. The ns-electron pulses (~$10^4$ electrons per pulse) used for the real-time imaging in our 4D-EM are too weak to generate any bubbles in the liquid cell.

## Supplementary Text

### Physical model analysis

For a Brownian particle in the presence of an external force in the liquid cell, its one-dimensional translational dynamics is governed by the Langevin equation[2]

$$\frac{d^2}{dt^2}x + \gamma \frac{d}{dt}x = A_{ext}(t) \qquad (1)$$

where $\gamma$ is the damping factor due to surrounding drag including both the fluid friction force and the interaction from the substrate, $A_{ext}(t)$ is the acceleration executed by an external fluctuation force. The time dependent displacement $x(t)$ is given by

$$x(t) = \frac{1}{\gamma}\int_0^t d\tau \left(1 - e^{-\gamma(t-\tau)}\right) A_{ext}(\tau) \qquad (2)$$

The random force upon the gold NP under repetitive fs-laser pulses excitation is assumed to be caused by steam nanobubbles (NBs) induced by laser pulse heating of the NP over water boiling temperature. The laser pulse heats up the particle in hundreds of picoseconds due to the surface plasmon enhanced strong optical absorption at the wavelength of 520 nm[3]. Such a strong local photothermal effect induces rapid and repetitive nucleation, expansion and collapse of NBs near the particle surface[4], resulting in a sequence of repetitive impulsive random forces which drive the superfast diffusion of the particle. For simplicity, we assume that the acceleration $A_{ext}(t)$ arising from the NBs consists of a comb of very short rectangular-shaped impulses induced by the repeated laser pulses heating, with a duration of $\tau_0$ and a time interval of $T_p$ between the periodic impulses (repetition rate is $1/T_p$). For the k-th impulse one has

$$A_{ext}(t) = A_k(0)\left(H(t-kT_p) - H(t-\tau_0-kT_p)\right) \tag{3}$$

where $H(t)$ is a Heaviside step function. Thus the Langevin equation (1) can be recast as

$$\frac{d^2}{dt^2}x + \gamma \frac{d}{dt}x = \sum_{k=0} A_k(0)\left(H(t-kT_p) - H(t-\tau_0-kT_p)\right) \tag{4}$$

For simplicity, one can assume $x(0)=0$ initially, but the velocity $u_0 = \dot{x}(0) \neq 0$. Assuming the damping constant is independent of liquid temperature and resolving the Langevin equation (4), we can obtain

$$x(t) = \frac{u_0}{\gamma}\left(1-e^{-\gamma t}\right) + \sum_{k=0} \frac{A_k(0)}{\gamma}\int_0^t d\tau \left(1-e^{-\gamma(t-\tau)}\right)\left(H(\tau-kT_p) - H(\tau-\tau_0-kT_p)\right) \tag{5}$$

For $0 \leq t < \tau_0$,

$$x(t) = \frac{u_0}{\gamma}\left(1-e^{-\gamma t}\right) + \frac{A_0(0)}{\gamma}t - \frac{A_0(0)}{\gamma^2}e^{-\gamma t}\left(e^{\gamma t}-1\right) \tag{6}$$

and for $\tau_0 \leq t < T_p$,

$$x(t) = \frac{u_0}{\gamma}\left(1-e^{-\gamma t}\right) + \frac{A_0(0)}{\gamma}\tau_0 - \frac{A_0(0)}{\gamma^2}e^{-\gamma t}\left(e^{\gamma \tau_0}-1\right) \tag{7}$$

When $T_p \leq t < T_p + \tau_0$,

$$x(t) = \frac{u_0}{\gamma}\left(1-e^{-\gamma t}\right) + \frac{A_0(0)}{\gamma}\tau_0 + \frac{A_1(0)}{\gamma}(t-T_p) - \frac{A_0(0)}{\gamma^2}e^{-\gamma t}\left(e^{\gamma \tau_0}-1\right) - \frac{A_1(0)}{\gamma^2}e^{-\gamma t}\left(e^{\gamma t}-e^{\gamma T_p}\right) \tag{8}$$

and for $\tau_0 + T_p \leq t < 2T_p$,

$$x(t) = \frac{u_0}{\gamma}\left(1-e^{-\gamma t}\right) + \frac{A_0(0)+A_1(0)}{\gamma}\tau_0 - \frac{A_0(0)}{\gamma^2}e^{-\gamma t}\left(e^{\gamma \tau_0}-1\right) - \frac{A_1(0)}{\gamma^2}e^{-\gamma t}\left(e^{\gamma(T_p+\tau_0)}-e^{\gamma T_p}\right) \tag{9}$$

Similarly, for $(N-1)T_p \leq t < (N-1)T_p + \tau_0$,

$$x(t) = \frac{u_0}{\gamma}\left(1-e^{-\gamma t}\right) + \sum_{k=0}^{N-2}\frac{A_k(0)}{\gamma}\tau_0 + \frac{A_{N-1}(0)}{\gamma}(t-T_p)$$

$$-\sum_{k=0}^{N-2}\frac{A_k(0)}{\gamma^2}e^{-\gamma t}\left(e^{\gamma(kT_p+\tau_0)} - e^{\gamma kT_p}\right) - \frac{A_{N-1}(0)}{\gamma^2}e^{-\gamma t}\left(e^{\gamma kt} - e^{\gamma kT_p}\right)$$

(10)

and for $(N-1)T_p + \tau_0 \leq t < NT_p$,

$$x(t) = \frac{u_0}{\gamma}\left(1-e^{-\gamma t}\right) + \sum_{k=0}^{N-1}\frac{A_k(0)}{\gamma}\tau_0 - \sum_{k=0}^{N-1}\frac{A_k(0)}{\gamma^2}e^{-\gamma t}\left(e^{\gamma(kT_p+\tau_0)} - e^{\gamma kT_p}\right)$$

(11)

We could calculate the time-dependent mean square displacement (MSD) according to the above equations. Since there is no correlation between the initial velocity and the impulsive forces, one has $\langle u_0 A_k(0)\rangle = 0$. In addition, the direction of the impulsive force induced by different shots has no correlation either, thus $\langle A_k(0)A_j(0)\rangle = \delta_{kj} A_{ave}^2$, where $A_{ave}$ represents an average value of the comb of random impulses. $A_{ave}$ is reasonably assumed to be proportional to the laser fluence. Because $\langle A_k(0)\rangle = 0$, the ensemble average for the displacement reads $\langle x(t)\rangle = \frac{u_0}{\gamma}\left(1-e^{-\gamma t}\right)$.

Therefore, at the long time scale such as $t = NT_p$, we can obtain the MSD as

$$MSD = \langle x^2(t)\rangle - \langle x(t)\rangle^2 = \sum_{k=0}^{N-1}\frac{A_{ave}^2}{\gamma^4}\left[\gamma\tau_0 - e^{-\gamma NT_p}e^{\gamma kT_p}\left(e^{\gamma\tau_0}-1\right)\right]^2$$

$$= \frac{A_{ave}^2}{\gamma^4}\left[\frac{t}{T_p}\gamma^2\tau_0^2 - 2\gamma\tau_0\left(e^{\gamma\tau_0}-1\right)\frac{1-e^{-\gamma t}}{e^{\gamma T_p}-1} + \left(e^{\gamma\tau_0}-1\right)^2\frac{1-e^{-2\gamma t}}{e^{2\gamma T_p}-1}\right]$$

(12)

For $t \gg 1/\gamma$, the MSD reads

$$MSD \approx \frac{A_{ave}^2\tau_0^2}{\gamma^2}\left(\frac{1}{e^{2\gamma T_p}-1} - \frac{2}{e^{\gamma T_p}-1}\right) + \frac{t}{T_p}\frac{A_{ave}^2\tau_0^2}{\gamma^2}$$

(13)

Since the first term is negligibly small when $\gamma T_p \gg 1$, the MSD approaches a linear time dependence on the long time limit as given in the following which is the key formula derived in this work

$$MSD \approx \frac{t}{T_p} \frac{A_{ave}^2 \tau_0^2}{\gamma^2} \quad (14)$$

Similarly, for a two-dimensional translational dynamics of an active Brownian particle, one can get $MSD \approx 2\frac{t}{T_p}\frac{A_{ave}^2 \tau_0^2}{\gamma^2}$. In analog with Einstein's linear time dependence law for a conventional two-dimensional passive Brownian motion, i.e. $MSD = 4Dt$, we can define an effective two-dimensional diffusion constant for the photon-activated Brownian particle in as $D = \frac{A_{ave}^2 \tau_0^2}{2T_p \gamma^2}$. The predicted effective diffusion constant $D$ has a quadratic dependence on the strength of the impulsive acceleration, which increases with the laser fluence.